\documentclass[12pt,preprint]{aastex}
\usepackage{natbib}
\usepackage[normalem]{ulem}

\begin{document}

\title{A Density Dependence for Protostellar Luminosity in Class I
Sources: Collaborative Accretion}

\author{
Bruce G. Elmegreen\altaffilmark{1}, Rachel Hurst\altaffilmark{2}, and Xavier
Koenig\altaffilmark{3} } \altaffiltext{1}{IBM Research Division, T.J. Watson
Research Center, Yorktown Hts., NY 10598, bge@us.ibm.com}
\altaffiltext{2}{Scarsdale High School, 1057 White Plains Rd, Scarsdale, NY
10583} \altaffiltext{3}{Department of Astronomy, Yale University, New Haven, CT
06520, USA}

\begin{abstract}
Class I protostars in three high-mass star-forming regions are found to have
correlations among the local projected density of other Class I protostars, the
summed flux from these other protostars, and the protostellar luminosity in the
WISE $22\mu$m band. Brighter Class I sources form in higher-density and
higher-flux regions, while low luminosity sources form anywhere. These
correlations depend slightly on the number of neighbors considered (from 2 to
20) and could include a size-of-sample effect from the initial mass function
(i.e., larger numbers include rarer and more massive stars). Luminosities seem
to vary by neighborhood with nearby protostars having values proportional to
each other and higher density regions having higher values. If Class I
luminosity is partially related to the accretion rate, then this luminosity
correlation is consistent with the competitive accretion model, although it is
more collaborative than competitive.  The correlation is also consistent with
primordial mass segregation, and could explain why the stellar initial mass
function resembles the dense core mass function even when cores form multiple
stars.
\end{abstract}


\section{Introduction}

Stars in a cluster often segregate by mass in the sense that the more massive
stars are located closer to the center \citep[e.g.,][]{degrijs02,er13}. The same
is true for loose stellar groups \citep{kirk11}. This is a natural result of
mutual stellar scattering over a relaxation time \citep{spitzer40}, but often
mass segregation appears long before this \citep{hillenbrand98,
bonnell98,gouliermis04,gennaro11}, leading to questions about its origin and
whether it reveals processes relevant to the stellar initial mass function
(IMF).

One possible cause of mass segregation is that dynamical friction between
orbiting protostars or prestellar clumps and gas in the embedded cluster phase
causes a greater drag on the more massive objects, bringing them closer to the
center \citep{gorti95, pelu12,indulekha13}. Another possibility is that initial
subclusters relax quickly and preserve their mass segregation upon merging into
the final cluster \citep{mcmillan07,moeckel09,allison10,maschberger10,yu11},
although it may take several orbit times for the cluster to show it
\citep{bate09}. A third possibility is that stellar mass increases with local
density and temperature at the time of formation \citep{larson82,bonnell01}.

Additional processes leading to mass segregation occur in simulations of
collapsing clouds \citep{bonnell97,bonnell06}.  Supersonic turbulence forms
filaments that channel the collapse into dense cores; the
accretion rate is larger in more centralized cores, so they produce more massive
stars if they do not fragment \citep{girichidis12}. The sharing of gas by
protostars in a core is known as competitive accretion
\citep[e.g.,][]{zinnecker82}.

Coagulation of pre-stellar clumps or protostars also produces more massive stars
in central regions \citep{bonnell05,moeckel11}.  Coagulation could be important
in the most massive clusters \citep[e.g.,][]{port02}. When it operates, the most
massive star tends to run away from the others in mass, taking most of the low
mass objects in its vicinity \citep{moeckel11}. The result is a disjoint IMF,
with a gap between the most massive star and the next most massive star in the
same core. This could lead to the appearance of isolated massive star formation.
Some massive stars apparently do form in relative isolation, which means they do
not have the normal proportion of lower mass stars around them
\citep[e.g.][]{dewit05,bressert12,oey13}. Isolated massive star formation may be
viewed as the antithesis of competitive accretion \citep{mckee03,zhang13},
although stochastic variations and coagulation could produce disjoint or
top-heavy IMFs in a competitive accretion environment when the number of stars
is small.

Mass segregation has recently been measured using a minimum spanning tree. The
mean segment length of the minimum spanning tree for each mass interval is found
to get longer for lower masses, down to some particular mass
\citep{allison09,olczak11,pang13,goldman13}.  If the dynamical relaxation time
at the limiting mass is comparable to the cluster age, then there has been
enough time for this cluster to segregate by stellar scattering. The correlation
between spanning tree length and mass is not seen for all clusters, however
\citep{parker12}. The method was applied to young stellar objects by
\cite{gutermuth09}.

Here we measure the projected density as a function of $22\mu$m luminosity for
Class I objects in several nearby star-forming regions, using data from the {\it
Wide Field Infrared Survey Explorer} (WISE) all sky survey \citep{wright10}.
With typical ages of $<10^6$ yrs \citep{prato09,enoch09} and velocities
comparable to or less than the turbulent speed of cluster-forming cores
\citep[$\sim1$ km s$^{-1}$, ][]{offner09}, most Class I stars are too young to
have moved significantly relative to other stars, so their locations should be a
good indicator of where they formed \citep[however, see][]{heiderman10}. Their
luminosities are dominated by radiation in the far-infrared, which allows their
discovery even in highly obscured young clusters. Luminosity is also related to
stellar mass through intrinsic and accretion components, although the latter
could be variable \citep{evans09}.

A similar study by \cite{kryukova12} showed that higher luminosity protostars
form in regions with higher densities of other protostars, as measured by the
distance to the fourth nearest protostar.  This appears to be evidence for
primordial mass segregation. Krukova et al. also found that regions with more
massive stars have more massive protostars, suggesting a regional dependence for
the IMF, or a common environment for protostars in a region \citep{offner11}. We
also find here that the most luminous protostars occur in the densest regions.
Because denser regions often have larger sample sizes, and larger samples can
have more massive stars just by stochastic effects, we check whether the
luminosity-density correlation contains a size-of-sample effect. The results
suggest that it may: the brightest source in a region satisfies $L_{\rm
22\mu{\rm m}}\sim6500 n_{\rm Class \;I}^{2}N^{0.8}\;L_\odot$, for $22\mu$m
luminosity $L_{\rm 22\mu{\rm m}}$, projected density $n_{\rm Class\;I}$ in Class
I sources pc$^{-2}$ and number of sources considered, $N$.  The Salpeter IMF has
a maximum likely mass $\propto N^{0.74}$.

In addition, we find a correlation between the $22\mu$m magnitude of a Class I
source and the summed $22\mu$m flux of other nearby Class I $22\mu$m sources.
This correlation seems to suggest that accretion rates, which partially
determine the Class I luminosities, vary from region to region with higher or
lower rates applying to all of the stars simultaneously in the region. We refer
to this process as ``collaborative accretion'' to emphasize regional
co-variations of accretion rates. It could explain mass segregation like the
competitive accretion {model does,} but also predict that final stellar masses
in a regional subcluster vary together, possibly making the largest mass in the
local IMF proportional to the pre-stellar clump mass. With a proportionality
between all of the stellar masses in a clump and the clump mass, the final IMF
can resemble the initial clump mass function \citep{motte98} even when multiple
stars form in each clump.

\section{Data}

We searched for young stellar objects in the three richest star-forming regions
from the list of 17 regions studied by Koenig et al. (2012, see Table 1). Class
I sources were extracted for each region using color prescriptions in the
appendix of \cite{koenig12}, applied to the photometric source catalog extracted
by these authors from custom WISE mosaiced images, and then examined by eye to
eliminate spurious emission features. The selection criteria to find Class I
sources primarily rely on source colors in the three shortest WISE bands, with
supplementary criteria that use combined 2MASS and WISE to retrieve candidate
young stellar objects in cases of extremely bright backgrounds that mask the
band 3 detection. The criteria themselves are based on the colors of known young
stars in the Taurus region as cataloged by \cite{rebull10}.

Figure 1-3 (top left) show sky positions of the Class I sources using color to
denote relative magnitudes in the $22\mu$m passband (hereafter, apparent
magnitude will be denoted by ``w4'' and absolute magnitude by ``W4''). Objects
within 1 magnitude at $22\mu$m of the brightest source in the region are plotted
red, between 1 and 2 magnitudes are plotted green, and fainter than 2 magnitudes
from the brightest are in blue. The most luminous objects are generally
surrounded by the next most luminous, with the low luminosity sources all around.

Some candidate Class I sources could be unresolved clusters. A preliminary look
suggests that about half of the sources visible in the Spitzer MIPS $24\mu$ band
at twice the angular resolution of WISE are single sources in both, and the rest
could be multiple. Still, an analysis of pure MIPS $24\mu$m detections in the W5
region gives about the same correlations as we report here using WISE band 4.
Thus the luminosities of unresolved clusters may have the same density and
near-neighbor properties as the luminosities of individual protostars. Such
similarity may follow from proportional cluster and prototellar accretion rates.

We use $22\mu$m as a measure of bolometric luminosity for our sources because
that and the $11.6\mu$m band each contain about half of the total luminosity in
the near-infrared range from J-band to $22\mu$m. \cite{dunham13} presented
bolometric luminosities using near-IR through sub-millimeter photometry for
protostars in the c2d survey of nearby star forming clouds. His MIPS $24\mu$m
luminosities scale well with the bolometric luminosity for high temperature
sources ($T>70$ K), and these are the sources we consider with our Class I
selection.

\section{Density and Neighborhood Flux around Class I Sources}

To quantify the visual impression that object luminosity clusters
hierarchically, we measured the average density of the $N$ nearest Class I
sources around each Class I source. The top right panels in Figures 1-3 show, as
points, the density of neighbors around each Class I source, $n_i$, measured out
to a fixed number, $N=10$, of neighbors, and plotted versus the absolute
magnitude of source $i$ in the $22\mu$m passband, $W4_i$. The density for source
$i$ is taken to be $n_i=N/(\pi R_{N,i}^2)$ for projected distance $R_{N,i}$ in
parsecs to the $N^{th}$ neighbor around source $i$; the density does not include
the source itself.  Faint Class I sources have a wide variety of densities of
other Class I sources around them, but bright Class I sources have only high
densities of other Class I sources around them.

The lower limit to the distribution of points in these figures may be viewed as
the minimum neighbor density for the occurrence of a Class I source of magnitude
$W4$. To calculate the functional form of this lower limit, we bin the
magnitudes into intervals of 1 magnitude, and average together the logarithms of
the three lowest densities in each interval. Then we fit a linear relationship
between this average, $n_{\rm limit}$ and the magnitude, $W4$,
\begin{equation}
\log n_{\rm limit}= A + B\times W4
\end{equation}
The fit depends on the number of neighbors, $N$. The crooked solid lines in the
top right panels show these minimum log-densities versus $W4$; the dashed lines
with the same colors are the fits. The different colors are for different $N$:
blue for $N=2$, green for $N=5$, red for $N=10$ and black for $N=20$. The points
themselves are for $N=10$, as mentioned above.

The bottom right panels in Figures 1-3 show the total flux at each source $i$
from the $N$ nearest neighbors, versus $W4_i$. The flux is calculated from the
sum of the fluxes from each of the $N$ sources inside that neighborhood. A
convenient measure of flux involves the absolute $W4$ magnitude as
representative of stellar luminosity, and the projected distance $R_{ij}$ in
parsecs between source $i$ and the $j^{th}$ nearest Class I source:
\begin{equation}
F_i=\Sigma_{j=1}^{N} 10^{-0.4{\rm W4}_j}/R_{ij}^2 \label{flux}
\end{equation}
The average of the three lowest (log) fluxes is shown versus the magnitude as
crooked lines, one for each value of $N$. The dashed lines of the same color are
the linear fits.

The units of projected density in the figures are Class I sources per square
parsec. The units of flux are inverse square parsecs and can be converted to
energy fluxes as follows. First, the tabulated magnitudes in the WISE database,
e.g., $w4$, are apparent magnitudes for flux density, which can be converted to
flux per unit frequency at the Earth using the
equation\footnote{http://wise2.ipac.caltech.edu/docs/release/prelim/expsup/sec4\_3g.html\#WISEZMA}
\begin{equation}
F_{\rm Earth}=8.363\times10^{-0.4{\rm w4}}\;{\rm  Jy}.
\end{equation}
In the Figures, the plotted quantity for flux density at $22\mu$m is what would
be measured at the location of a Class I source from some other Class I source
nearby. In an analogous fashion, we write this neighborhood flux as $L/(4\pi
R_{\rm ij}^2)$ for luminosity $L=4\pi D^2\Delta \nu F_{\rm Earth}$, where
$R_{\rm ij}$ is the projected distance between sources $i$ and $j$, $D$ is the
distance from the Earth, and $\Delta \nu=2.4961\times10^{12}$ Hz is the $22\mu$m
bandwidth. Converting to absolute magnitude and using $D=10$ pc in that case,
and dividing out $\Delta\nu$, we obtain for flux density
\begin{equation}
F_{\rm i}=836.3\times \Sigma_{j=1}^{N} 10^{-0.4{\rm W4}_j}/R_{\rm ij}^2\;{\rm Jy}
\end{equation}
for $R_{\rm ij}$ in pc as in the figures. Thus the dimensional flux density at
source $i$ is $836.3$ Jy times the plotted flux in the Figures.

Figure 4 gives the resultant values of the fitting coefficients for $N=2$, 5, 10
and 20 for the minimum density and minimum flux correlations. The slope, $B$, is
written as the derivative, $d\log{\rm density}/d\;{\rm W4}$ or $d \log {\rm
Flux}/d\;{\rm W4}$, and the intercept, $A$, is written as the density or flux at
an absolute magnitude of $W4=-10$. We pick these values of $N$ because larger
$N$ give neighborhoods that are approaching the size of the region, making the
densities unrealistic for objects near the edge of the field. Error bars in
Figure 4 are 60\% uncertainty limits using student-t statistics.

The slope of the correlation between minimum density of class I sources, $n_{\rm
Class\;I}$ and $W4$ (Fig. 4a) increases slightly with $N$ and has an average
value of $\sim-0.2$ (formally, $-0.19\pm0.06$). Considering the logarithmic
relation between magnitude and luminosity $L_{\rm 22\mu}$, this density slope
implies that $L_{\rm 22\mu{\rm m}}\propto n_{\rm Class\;I}^2$. The intercept of
this correlation (Fig. 4b) decreases slightly with $N$ but averages $\sim-1.5$
(formally, $-1.48\pm0.19$). Considering the calibration above, this gives
$L_{\rm 22\mu{\rm m}}\sim6500 n_{\rm Class \;I}^{2}\;L_\odot$ at the minimum
density for each $W4$ magnitude.

The decreasing trend in Figure 4b has an average slope of $-0.42\pm0.18$. The
trend may be related to a size-of-sample effect, which gives a slope $-0.37$ as
shown by the black line (arbitrarily shifted vertically). That is, for a
Salpeter IMF where $N(M)\propto M^{-2.35}$, the number of stars $N$ in a sample
scales with the maximum mass as $M_{\rm max}^{1.35}$, on average. If final
stellar mass is proportional to Class I luminosity (a big assumption), and the
maximum luminosity $L$ is also $\propto n^2$ for density $n$ from the top-left
panel, then overall $L\propto n^2N^{1/1.35}$ in a probabilistic sense. This
means that at fixed $L$, $n\propto N^{-1/2.7}$ and that is the black line.

The minimum flux in the neighborhood of a Class I source depends on the number
of neighbors (Fig. 4d) because this flux is the sum over the fluxes from each
neighbor considered. The correlation itself varies slightly with the region.

The bottom left panels in Figures 1-3 plot the projected separations in parsecs
between all Class I sources $i$ and their $N=10$ nearest neighbors, $j$, versus
the absolute value of the magnitude difference between sources $i$ and $j$.
Symbol color indicates relative brightness for source $i$ as in the top left.
The figure shows a triangular distribution of points for each color, which
implies that closer neighbors have more similar magnitudes in each brightness
interval. This differential correlation is consistent with the positive
correlation between neighborhood flux and source luminosity (Figs. 1-3 lower
right). What it means is that nearby Class I sources in a region have neighbors
with luminosities proportional to each other.  Figure 12 in Kryukova et al.
(2012) has a similar triangular distribution but it plots magnitude versus
distance to the 4th nearest source. Their plot shows that brighter sources occur
in denser regions, which we also show here in our top-right panels of Figures
1-3. The two lower panels in Figures 1-3 show that neighbor fluxes correlate
with each other over a wide range of luminosities. The top part of the triangle
is filled in because different subclusters that are widely separated can have
equally bright sources.

\section{Discussion}

Figures 1-3 show luminosity segregation already at the young age of a Class I
source, confirming a similar result by \cite{kryukova12}.  The most luminous
sources are primarily in the densest regions, while lower luminosity sources
occur anywhere in the field. The most luminous sources also occur in regions
with the strongest $22\mu$m fluxes from other Class I sources nearby, while the
lower luminosity sources can occur anywhere. Both correlations could be the
result of primordial mass segregation if luminosity correlates with mass. In
addition, sources near a Class I source have similar luminosities, which implies
that protostellar accretion rates may be a property of neighborhoods to the
extent that luminosity also correlates with accretion rate. Mass segregation
logically follows if accretion rates are higher at higher subcluster densities.
This result raises the interesting possibility that accretion may be {\it
collaborative}, rather than competitive, which means there is mutually enhanced
accretion among protostars that lie near each other in dense regions.

The maximum $22\mu$m luminosity in a region is found to scale with the square of
the density of Class I sources, and with a power $\sim0.84$ of the number of
sources considered  (i.e., double the measured power for the density-number
relation).  The size-of-sample effect for the IMF predicts approximately this
latter power.

Collaborative accretion in subclusters may solve a puzzle related to the
similarity between the core mass function and the star mass function. This
similarity makes sense if single cores form single stars with proportional mass
\citep[e.g.,][]{alves11}, but it does not make sense if cores form multiple
stars with the standard IMF in each. However, if the mass of the core scales
with the accretion rate onto the core, and the masses of all of the stars in the
core scale with this accretion rate too, then all of the stellar masses in that
core become proportional to core mass.

We are grateful to D.T. Leisawitz for discussions about the use of WISE data.
R.H. is grateful to Beth Schoenbrun for helpful discussions. XK acknowledges
support from NASA ROSES ADAP grant NNX13AF07G. Helpful comments by the referee
are appreciated. This publication makes use of data products from the Wide-field
Infrared Survey Explorer, which is a joint project of the University of
California, Los Angeles, and the Jet Propulsion Laboratory/California Institute
of Technology, funded by the National Aeronautics and Space Administration.

\clearpage

\begin{deluxetable}{lcccccc}
\tabletypesize{\scriptsize} \tablecaption{Star Forming Regions} \tablehead{
\colhead{Name} & \colhead{RA (deg)} & \colhead{DEC (deg)} & \colhead{Dist.
(kpc)} & \colhead{$22\mu$m FWHM (pc)} & \colhead{Number of Class I} &
\colhead{Class I with $22\mu$m fluxes\tablenotemark{a}} } \startdata
W3, W4, W5 & 34-46      &  59-63    & 2.1       &  0.12 & 356 & 184   \\
IC 1396    & 320-330    &  55-60    & 0.9       &  0.05 & 245 & 164   \\
NGC 2175   & 91.6-93.2  &  20-21    & 2.6       &  0.15 & 169 & 86
\enddata
\tablenotetext{a}{The figures, densities, and analyses in this paper consider
only these Class I sources with $22\mu$m fluxes.} \label{tab1}
\end{deluxetable}

\clearpage
\begin{figure}
\includegraphics[width=3.5in]{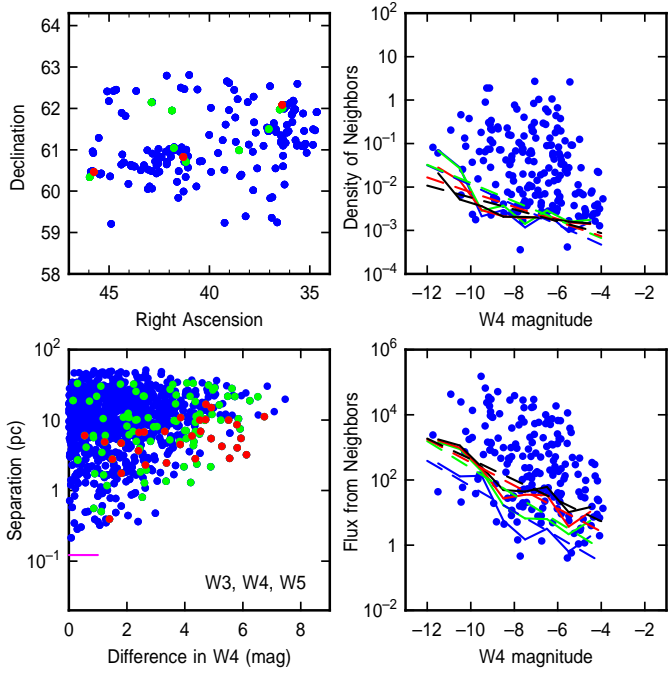}
\caption{Results for W3, W4 and W5: (top left) Sky view of Class I sources having
$22\mu$m fluxes, with color
indicating luminosity range: sources within 1 magnitude of the brightest source
are red, between 1 and 2 magnitudes are green, and 2 magnitudes or more fainter
than the brightest are blue. (top right) Projected density in pc$^{-2}$ versus
absolute source magnitude in the $22\mu$m band of WISE. Points are for $N=10$
neighbors considered in the density. Crooked lines connect the average of 3 lowest
densities in each magnitude interval; dashed lines are linear fits to the crooked
lines. (bottom right) Flux at the position of a protostar, measured in pc$^{-2}$, versus
the absolute magnitude in the $22\mu$m band. Points are for $N=10$;
crooked and dashed lines represent
averages of the lowest 3 fluxes in each magnitude interval for various $N$.
(bottom left) The projected
separation, in pc, versus the difference in absolute
magnitude, for all sources $j$ within $N=10$ neighbors of each source $i$. Colors
mean the same as in the top left panel. The $12^{\prime\prime}$ FWHM at $22\mu$m is
indicated by a line on the y-axis.}
\end{figure}

\begin{figure}
\includegraphics[width=3.5in]{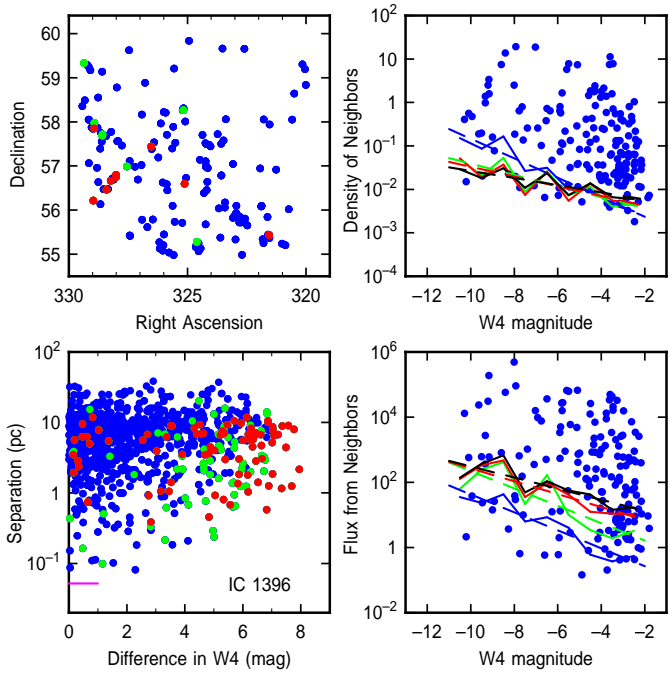}
\caption{Results for IC 1396, as in figure 1.}
\end{figure}

\begin{figure}
\includegraphics[width=3.5in]{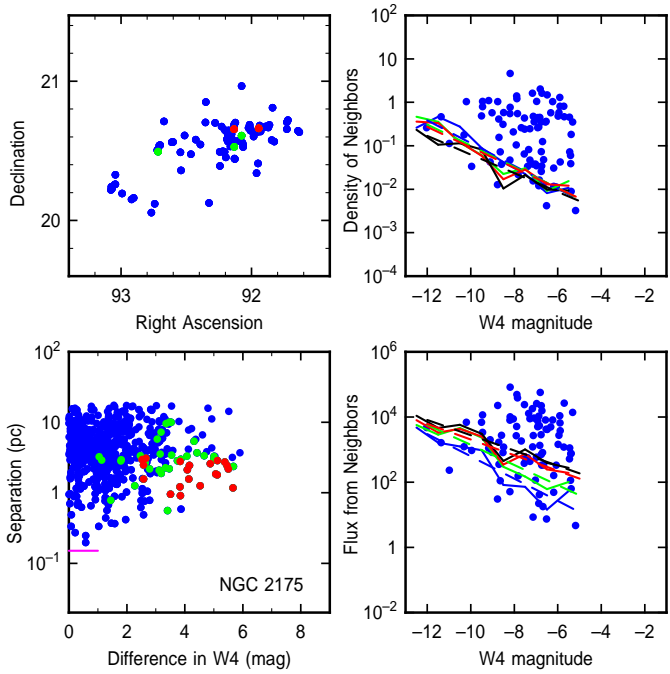}
\caption{Results for NGC 2175, as in figure 1.}
\end{figure}

\begin{figure}
\centering
\includegraphics[width=6.5in]{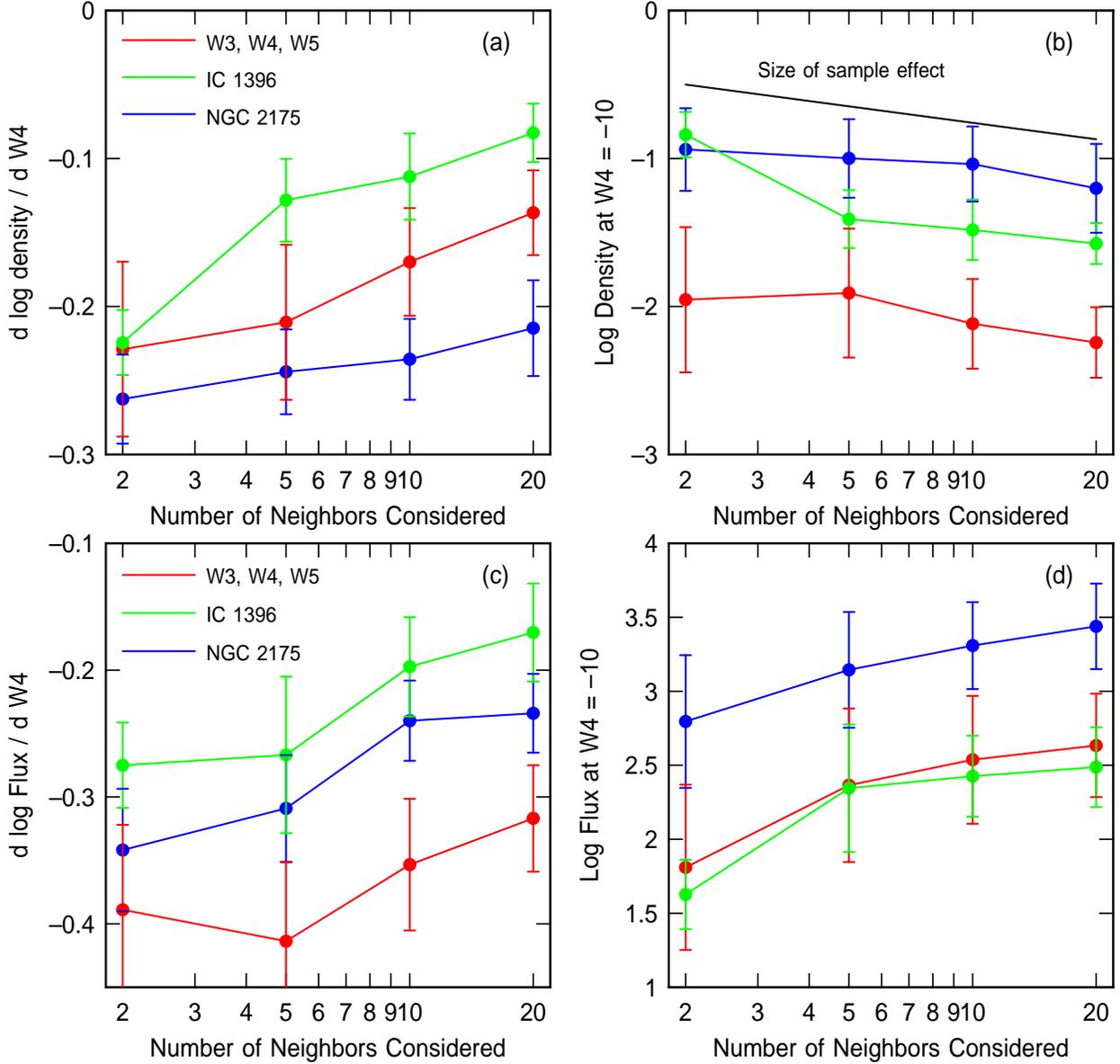}
\caption{Slope (left) and intercept (right) of the fitted linear correlation between the minimum flux at a
protostar and the protostellar absolute magnitude (bottom two panels), and the same
for the minimum density (top two panels). The intercept is written as the flux or density
(both in units of pc$^{-2}$) for a source with an absolute magnitude of $-10$ in the
$22\mu$m band.  The abscissa in each case is the number of neighboring sources used to
determine the local flux or density. The purpose of this figure is to look for
size-of-sample effects. The minimum flux increases with $N$ in a trivial way because
larger $N$ corresponds to a larger sum of neighbor fluxes. The minimum density decreases
slightly with $N$, somewhat parallel to the black line, which is the predicted size-of-sample
effect for a Salpeter IMF given the average scaling between density and luminosity in figure (a)
and assuming luminosity is proportional to mass.
}\label{fig1}\end{figure}

\end{document}